\documentclass[prb,twocolumn,superscriptaddress,floatfix]{revtex4}
\usepackage{hyperref}
\usepackage{graphicx}
\usepackage{amsmath}
\usepackage[dvips]{epsfig}
\newcommand{\beq}{\begin{equation}}
\newcommand{\eeq}{\end{equation}}
\newcommand{\beqa}{\begin{eqnarray}}
\newcommand{\eeqa}{\end{eqnarray}}
\newcommand{\ba}{\begin{array}}
\newcommand{\ea}{\end{array}}

\begin{document}

\title{Vortices in quantum droplets of heteronuclear Bose mixtures}

\author{Matteo Caldara}
\affiliation{Dipartimento di Fisica e Astronomia ``Galileo Galilei''
, Universit\`a di Padova, via Marzolo 8, 35122 Padova, Italy.}
\affiliation{International School for Advanced Studies (SISSA), Via Bonomea 265, 34136 Trieste, Italy.}
\author{Francesco Ancilotto}
\affiliation{Dipartimento di Fisica e Astronomia ``Galileo Galilei''
and CNISM, Universit\`a di Padova, via Marzolo 8, 35122 Padova, Italy.}
\affiliation{CNR-IOM Democritos, via Bonomea, 265 - 34136 Trieste, Italy.}

\begin{abstract}
\noindent We have theoretically investigated 
the structure of spinning self-bound droplets made of $^{41}$K-$^{87}$Rb 
Bose mixture by solving
the Gross-Pitaevskii equation including 
beyond-mean-field correction in the Lee-Huang-Yang form.
The structure and energetics of vortex formation in 
the self-bound mixture have been elucidated,
showing that the formation of linear vortices in the
heavier species is energetically favoured over 
other configurations. A fake (partially filled)
core develops as a consequence in the
other species, resulting in a hole which might be imaged in
experiments.
The interplay between 
vortices and capillary waves, which are the two ways angular momentum 
can be stored in a swirling superfluid, is studied in detail by
computing the relation between angular momentum and rotational
frequency. The results show intriguing similarities 
with the case of a prototypical superfluid, i.e. $^4$He droplets when 
set into rotation. 
A two-branches curve in the stability diagram, qualitatively similar 
to the one expected for classical (incompressible and viscous) rotating liquid droplets,
is obtained when vortices
are present in the droplets, 
while prolate (i.e. non axi-symmetric) shapes are only permitted in vortex-free droplets.

\end{abstract}
\date{\today}

\maketitle

\section{Introduction}

A new quantum state of matter has been predicted \cite{petrov2015quantum}, and
shortly after experimentally observed
\cite{cabrera2018quantum,semeghini2018self,derrico2019observation,NaRb2021} in
ultracold atomic gases made of a binary mixture of Bose atoms, where
the competition between
inter-species attractive interactions
and quantum fluctuations, which act as a repulsive interaction,
may result in the formation
of self-bound, ultradilute liquid droplets, with
typical densities being about eight orders of magnitude 
lower than those of the prototypical quantum fluid, i.e. liquid Helium, at room pressure.
Self-bound quantum liquid droplets have been predicted and observed
in dipolar Bose gases as well \cite{tanzi,Pfau,Ferlaino,PhysRevLett.126.025301},
with a similar stabilizing mechanism. 

Heteronuclear quantum droplets (QD in the following) in
a bosonic mixture of $^{41}$K and $^{87}$Rb
have been experimentally realized more recently
\cite{derrico2019observation}, and also in the $^{23}$Na-$^{87}$Rb mixture \cite{NaRb2021}. 
At variance with the 
largely studied homonuclear $^{39}$K-$^{39}$K mixture of $K$ atoms in two 
different hyperfine states,
the $^{41}$K and $^{87}$Rb mixture is characterized by longer lifetimes,
of the order of several tens of milliseconds,
i.e. more than a factor 10 larger than those
characterizing the $^{39}$K mixtures \cite{semeghini2018self}.
The longer lifetime is mainly a 
consequence of the smaller densities of the two components 
that result from the stronger intraspecies interactions. It follows that
the regime of interaction parameters for which self-bound droplets
form is such that three-body losses are expected to
be significantly reduced, 
at variance with the $^{39}$K-$^{39}$K mixture where much stronger three-body 
losses continuously drive the system out of equilibrium, eventually leading to 
the depletion of the droplet.
Longer lifetimes 
offer the possibility of investigating the collective
modes of the droplets, likely allowing 
the observation
of droplet self-evaporation \cite{modugno}.
Moreover, this will also favour 
the realization of larger droplets, 
characterized by a flat-top density profile encompassing 
a ``bulk" region with a nearly constant
saturation density, and a surface region whose width is determined by
the surface tension\cite{anci_poli}. 
The crossover from compressible (i.e. smaller droplets characterized by
an ``all surface", gaussian-like profile)
to incompressible (i.e. flat-top) quantum droplets, which is 
driven by the number of atoms, has been recently addressed experimentally
in $^{39}$K-$^{39}$K droplet collision experiments \cite{ferioli_collision},
and studied with numerical simulations \cite{ferioli_collision,viktor}.
However, a clean interpretation of the experimental results seems to be
hampered by the major role played in the $^{39}$K-$^{39}$K mixture 
by three-body loss factor, which is necessary 
in order to explain the experimental data \cite{ferioli_collision,viktor}, 
but whose actual value is affected by large uncertainties.
For this reason, the 
$^{41}$K-$^{87}$Rb mixture appears to be a better candidate for a
clear determination of the crossover.

Vortices are quantized topological 
excitations of superfluids, and have been extensively studied over the years
both in superfluid $^4$He \cite{donnelly,barenghi}
and in cold bosonic atoms \cite{fetter,Pit16,anderson}.
Although they should also
appear under suitable conditions in quantum droplets,
no experimental evidence of their existence has
been gathered so far. 
There exist, however, a number of theoretical 
papers adressing vorticity in quantum droplets made of Bose-Bose mixtures, 
mainly for 2-dimensional systems, which are briefly reviewed in the
following.

While vortices in quantum droplets made of dipolar condensates
are found to be always unstable
\cite{macri,malo_vort},
in binary Bose Einstein condensates (BEC) 
described by the Gross-Pitaevskii (GP) equation augmented 
by the beyond-mean-field correction in the Lee-Huang-Yang (LHY) form, 
they are found instead to be stable excitations when specific conditions are fulfilled
\cite{kartashov2019metastability,malomed,Mal18}.
Ground state and rotational properties of two-dimensional self-bound quantum droplets
made of a binary mixture of BECs
are studied in Ref. \onlinecite{examilioti2020ground}.
Several phases are found depending on the system parameters, 
including center of mass excitation, ghost vortices, and vortices with single 
and multiple quantization.
The metastability of clusters made of quantum droplets
is considered in Ref. \onlinecite{kartashov2019metastability} 
for a binary BEC in two dimensions,
leading to the formation of ring-shaped clusters, possibly hosting "supervortices".
Angular momentum-carrying droplets made of species-symmetric rotating binary BEC confined
in two dimensions \cite{tengstrand2019rotating} are found to be unstable in
free space and decay into fragments.
When stabilized in a weak harmonic trap,
and after switching off the trap potential, the
rotational ground state displays an array of few metastable
singly-quantized vortices, with significant distorsions
of the droplets shapes from the axi-symmetric configurations.
The effects of vorticity on the breathing modes of these droplets have
been addressed in Ref. \onlinecite{sturmer}. 
Two-dimensional droplets carrying vorticity 
are investigated in Ref. \onlinecite{kartashov2020structured}, where
axisymmetric QDs with heterosymmetric and heteromultipole structures, i.e.
with different vorticities in each component and/or different 
multipolarities (singly or doubly quantized) are studied.
The stability of vortical QDs was also studied in Refs. \onlinecite{Ota20,Mal18}.

To our knowledge, only 
two theoretical papers address vortical states 
in three-dimensional quantum droplets made of Bose mixtures.
In the first \cite{kartashov2018three},
droplets made of two-components superfluid Bose mixture under rotation 
are studied and
stationary states in the form of
vortex rings with embedded topological charges $m_1=m_2=1$ and $m_1=m_2=2$
of the two components are found for sizes larger than some critical values.
Droplets with
hidden vorticity, i.e. with topological charges $m_1= -m_2=1$ 
in the two components, are found instead to be always unstable
and split into fragments.
Equal scattering lengths $a$ of the contact interactions in both 
components are assumed in Ref. \onlinecite{kartashov2018three}, as well as equal masses for
the two species.

Singly and doubly quantized vortex states in a 3-dimensional 
droplet made of the heteronuclear $^{23}$Na-$^{87}$Rb bosonic mixture and 
of the homonuclear $^{39}$K-$^{39}$K mixture have been studied in Ref. \onlinecite{ancilotto2018self}.
A singly quantized vortex nucleated in both species is found to be stable and 
robust against quadrupolar deformation,
while the doubly quantized vortex eventually decays into pairs of
singly quantized vortices.
In both cases, the velocity field associated to the
angular momentum stored in the
doubly quantized vortex results in surface capillary 
waves that are responsible for the droplet distortion into a prolate shape and in 
the apparent rotation of the droplet as a whole.

We present here a study of the properties of spinning
three-dimensional droplets made of the $^{41}$K-$^{87}$Rb Bose mixture.
We compute several properties of vortices in such QDs and
focus on the relation between angular momentum, shape and
vorticity of quantum droplets.
These properties have been the subject of recent experimental and theoretical
studies on
the prototype superfluid Bosonic system, where spinning $^4$He nanodroplets
have been investigated in a series of experiments 
\cite{Gom14,Jon16,Ber17,Rup17,Lan18,Ges19,Oco20}.
Given the similarities between quantum droplets made of Bose mixtures
and liquid $^4$He droplets, as clarified in the following, 
we give here a brief account of recent
theoretical and experimental studies on spinning superfluid $^4$He
nanodroplets. 

We first recall here (the following discussion is partly taken from 
Ref.\onlinecite{Anc21}) that 
the macroscopic behaviour of a superfluid
at zero temperature is described by the equations of irrotational hydrodynamics,
from which the moment of inertia along the $z$-axis can be calculated as\cite{Boh75,Pit16}
$\Theta_{\rm irr} = \varepsilon^2 \, \Theta _{\rm rig}$, where
\begin{equation}
\varepsilon =\frac{\langle y^2-x^2 \rangle}{\langle y^2+x^2 \rangle}
\label{eq61}
\end{equation}
and
$\Theta _{\rm rig}=N m \langle x^2+y^2\rangle$ is the rigid-body moment of inertia, $N$ is the
number of atoms in the droplet and $m$ the atomic mass,
showing that in a superfluid the value of the moment of inertia
is smaller than the value for a rigid-body system.
In particular, the above relation shows that for axisymmetric 
(i.e. oblate) systems, where $ \langle x^2 \rangle =\langle y^2 \rangle$,
the angular momentum of the superfluid along the $z$-axis vanishes,
$\langle \hat{L}_z \rangle =\Theta_{\rm irr} \, \omega =0$,
for any value of the rotational frequency. 
Therefore oblate samples of a superfluid
cannot spin, whereas prolate (non axisymmetric)
configurations can, and the resulting angular momentum 
$L_{\rm cap} =  \Theta_{\rm irr} \, \omega$ is associated with
the presence of capillary waves (see, for instance, Ref. \onlinecite{Whi98}).
Quantized vortex lines represent the other well known 
mechanism with which angular momentum can be stored
in spinning superfluids.
In general, capillary waves and vortices may coexist in spinning
droplets, as shown 
by experiments and theory for the $^4$He case\cite{Oco20,Anc21}.
In the experiments \cite{Gom14,Jon16,Ber17,Rup17,Lan18,Ges19,Oco20},
where $^4$He liquid droplets may acquire angular momentum 
during the passage of the fluid through the nozzle of the
molecular beam apparatus, and where 
they were able to reconstruct 
images of the rotating $^4$He droplets,
{\it oblate} droplets were observed, which should be 
forbidden on quantum mechanical grounds.
Thus the only possible explanation for such 
experimental observation
is that these drops must contain quantized vortices which
can store most of the angular momentum of the droplet.

One striking outcome of these experiments was
the finding that spinning superfluid $^4$He droplets 
show unexpected similarities
with the behaviour of {\it classical}
incompressible viscous droplets, only subject
to surface tension and centrifugal forces\cite{Bro80,Hei06,But11,Bal15}.
It is precisely the presence of vortices in the droplet interior that confers on the
spinning droplet the appearance and the properties of a rotating,
classical viscous droplet, as also shown by 
density functional calculations \cite{Anc15,Anc18}. A well-known
example of this apparently classical behaviour of a rotating superfluid
is the macroscopic meniscus that develops, at the liquid-vapor interface,
in a rotating bucket filled with superfluid $^4$He above the critical
angular velocity required for vortex nucleation\cite{Osb50,donnelly}.
Similarities with the 
classical behaviour of rotating viscous droplets 
are also displayed, as shown in the following, by rotating QDs.

\section{Method}

The Gross-Pitaevskii energy functional for a Bose-Bose mixture,
including the Lee-Huang-Yang 
correction accounting for quantum fluctuations beyond mean-field
reads \cite{petrov2015quantum,ancilotto2018self}

\begin{widetext}
\begin{equation}
E = \sum_{i=1}^{2}\int d{\bf r} \, \left[{\hbar^2 \over 2m_i}|\nabla \psi_{i}({\bf r})|^2  
+ V_{i}({\bf r}) \rho_{i}({\bf r})\right]\,
+\frac{1}{2}\sum_{i,j=1}^{2}g_{ij}\int d{\bf r} \, \rho_{i}({\bf r})\rho_{j}({\bf r}) \,
+\int d{\bf r} \, {\cal E} _{\rm LHY}(\rho_1({\bf r}),\rho_2({\bf r}))\,
\label{eq:energyfun}
\end{equation}
\end{widetext}
where $V_i({\bf r})$ and $\rho_i({\bf r})=|\psi _i({\bf r})|^2$ 
represent the external potential and the (number) density of each component
($i=1$ for $^{41}$K, $i=2$ for $^{87}$Rb).The coupling constants are
$g_{11}=4\pi  a_{11} \hbar^2/m_1$, $g_{22}=4\pi a_{22} \hbar^2/m_2$, and $g_{12}=g_{21}=2\pi a_{12} \hbar^2/m_r$,
where $m_r =m_1m_2/(m_1+m_2)$ is the reduced mass.
The intra-species $s$-wave scattering lengths $a_{11}$ and $a_{22}$ are both positive, while the inter-species
one, $a_{12}$, is negative.
The scattering parameters describing the intraspecies repulsion
are fixed and their values are equal to $a_{11} = 65\,a_0$ \cite{derrico2007feshbach} 
and $a_{22} = 100.4\,a_0$ \cite{marte}. Notice that a slightly different value
for the K-K scattering length, $a_{11} = 62\, a_0$, has been used
more recently \cite{modugno}. 

The total number of bosons is $N=N_1+N_2$.
The number densities $\rho_1,\,\rho_2$ are normalized such
that $\int _V \rho_1({\bf r})\,{\rm d}{\bf r} =N_1$ and $\int _V \rho_2({\bf r})\,{\rm d}{\bf r} =N_2$.

The LHY correction is \cite{petrov2015quantum,ancilotto2018self}
\begin{align}
{\cal E} _{\rm LHY} &= {8\over 15 \pi^2} \left(\frac{m_1}{\hbar^2}\right)^{3/2}
\!\!\!\!\!\!(g_{11} \rho_1)^{5/2}
f\left(\frac{m_2}{m_1},\frac{g_{12}^2}{g_{11}g_{22}},\frac{g_{22}\,\rho_2}{g_{11}\,\rho_1}\right)
\nonumber
\\
&\equiv \mathcal{C} (g_{11}\rho_1)^{5/2}f(z,u,x).
\label{functional}
\end{align}
Here $f(z,u,x)>0$ is a dimensionless function, whose explicit expression
for $z\neq 1$ and $u=1$ can be found in Ref. \onlinecite{ancilotto2018self}.
Following Ref. \onlinecite{petrov2015quantum}, we consider this
function at the mean-field collapse $u=1$, i.e. $f(z,1,x)$.
We note that the actual expression for $f$ can be fitted very accurately with the
same functional form of the homonuclear case ($m_{1}=m_{2}$) \cite{Minardi}
\begin{equation}
f\left(z,1,x\right)\simeq \left(1+z^\alpha x\right)^{\beta}
\end{equation}
where $\alpha$ and $\beta$ are fitting parameters.
For the K-Rb mixture ($z=87/41$) we found $\alpha = 0.586$ and $\beta= 2.506$, which are
very close to the values $\alpha={3/5}$ and 
$\beta=5/2$ proposed in Ref. \onlinecite{Minardi}
under the assumption that $\alpha $ and $\beta $ are independent of the mass ratio $z$.

Minimization of the action associated to Eq.~(\ref{eq:energyfun}) leads to
the following Euler-Lagrange (EL) equations (\textit{generalized} GP equations) 

\begin{equation}
i \hbar {\partial \psi _i \over \partial t} =
\left[-{\hbar^2 \over 2m_i}\nabla ^2 + V_i + \mu_{i}(\rho_1,\rho_2) \right]\psi _i \equiv {\cal H}_i \psi _i
 \, ,
\label{eq:gpe}
\end{equation}
where
\begin{equation}
\mu_{i} = g_{ii}\rho_i+g_{ij}\rho_j+
\frac{\partial {\cal E}_{\rm LHY}}{\partial \rho_i}\quad \,( j \ne i) \,,
\label{eq:chempot}
\end{equation}
and
\begin{align}
\frac{\partial {\cal E} _{\rm LHY}}{\partial \rho_1} &=
\mathcal{C} g_{11}(g_{11}\rho_{1})^{3/2}\left({5\over 2}f-x{\partial f\over \partial x}\right)\,,
\\
\frac{\partial {\cal E} _{\rm LHY}}{\partial \rho_2} &=
\mathcal{C} g_{22}(g_{11}\rho_{1})^{3/2}{\partial f\over \partial x}\,.
\end{align}
where $\mathcal{C}$ is defined in Eq. \eqref{functional}.
The above equations are solved by mapping the system (densities,
wave functions, differential operators, etc.) on discrete equally
spaced cartesian grids. The differential operators 
are represented by a 13-point discretization.
We solve the above equations by propagating the wavefunctions $\psi_i $
in imaginary time, if stationary states are sought, or by
propagating them in real-time to simulate the dynamics of the system
starting from specified initial states.
The time-dependent equations have been
solved by using the Hamming's predictor-modifier-corrector method, 
initiated by a fourth-order
Runge-Kutta-Gill algorithm \cite{Anc17}.
The spatial mesh spacing and time step are chosen such that
during the time evolution excellent conservation
of the total energy of the system is guaranteed.

In order to deposit angular momentum in the droplet,
we have used an ``imprinting'' procedure \cite{Anc17} by starting the
imaginary time minimization from a flexible guess
for the  effective wave function 
$\psi_0(\mathbf{r})$ for a given species, namely a superposition of a quadrupolar 
capillary wave and $n_v$ vortex lines parallel to the $z$ axis,
\begin{equation}
\psi_0(\mathbf{r})=\rho_0^{1/2}(\mathbf{r})\,  e^{i \,\alpha x y} \,
\prod _{j=1}^{n_v} {(x-x_j)+i (y-y_j) \over \sqrt{(x-x_j)^2+(y-y_j)^2}}
\;
\label{eq57}
\end{equation}
Here, $\rho_0(\mathbf{r})$ is an arbitrary, vortex-free droplet
density, the complex phase $e^{i \alpha x y}$  
imprints a capillary wave with quadrupolar symmetry
around the $z$ axis, and the product term imprints 
a vortex array made of $n_v$ linear vortices\cite{Anc17},
where $(x_j, y_j)$ is the initial position of the $j^{th}$-vortex core.
The initial value of $\alpha$ and the vortex core positions are guessed, and
$\psi _0$ is optimized by iteratively solving the equations (\ref{eq:gpe}):
during the minimization process 
both the vortex core structure and positions, together with the droplet shape, change
to provide at convergence the lowest total energy configuration.

To study spinning droplets it is convenient
to work in the fixed-droplet frame of reference 
(corotating frame at angular velocity $\omega$), {\it i.e.} we consider
the functional
\begin{equation}
E' = E -  \omega \, \langle \hat{L}_z \rangle
\label{eq5}
\end{equation}
where $\hat{L}_z$ is the total angular momentum operator in the $z$-direction;
one looks for solutions of the EL equation resulting from the functional variation of  $E'$
\begin{equation}
\left\{{\cal H}_i \,- \omega \hat{L}_z\right\} \,\psi_i(\mathbf{r})  =  \,\mu _i\, \psi_i(\mathbf{r})
\label{eq7}
\end{equation}
where ${\cal H}_i$ ($i=1,2$) are defined in Eq. (\ref{eq:gpe}).

Two alternative strategies can be employed to solve the previous equations, 
{\it i.e.}, one can either (i) fix $\omega $ and
find the associated stationary configuration,
which will be characterized by some value of the angular momentum $L=\langle\hat{L}_z\rangle$
depending upon the chosen value of $\omega$, or (ii) solve it
by imposing a given value for $L$ and
iteratively find the associated value of $\omega$.
Classically, the fixed $\omega$ calculations correspond to
forced rotation conditions (``driven drops''), while the fixed $L$ calculations correspond
to torque free drops with an initially prescribed rotation (``isolated drops'').
Both methods will be used here, as it turns out 
that stable prolate configurations can only be found by using method (ii), {\it i.e.},
fixing the value of $L$ from the start \cite{Bro80,Hei06,But11}.
At variance, stable oblate configurations can be found either by
fixing $\omega$ or $L$.\cite{Bro80}

Working at fixed angular momentum requires to adjust iteratively the value of $\omega$:
there are efficient ways of doing this, such as
the Augmented Lagrangian method\cite{Gar21} (used here) which consists in evolving the system using the 
Hamiltonians
\begin{equation}
{\cal H}_i ^\prime = {\cal H}_i -[\omega -\mu _L (\langle\hat{L}_z\rangle-L)]\hat{L}_z
\end{equation}
and updating at each time step the angular velocity according to
\begin{equation}
\omega _{new} =\omega _{old} -\mu _L (\langle\hat{L}_z\rangle-L)
\end{equation}
where $\mu _L$ is a positive constant 
controlling the rate of convergence towards a state with the imposed value $L$ of 
the angular momentum.

\section{Results}

\subsection{Surface tension and healing length}

As discussed in the following Section III.E, 
results for rotating liquid droplets can be better interpreted in terms
of rescaled units of the rotational frequency and 
angular momentum, whose definitions require the knowledge of the 
surface tension of the system.
Moreover, the widths of the vortex cores in the quantum droplets  
are related to the healing lengths of the mixture.
For this reason, we report in the following
the calculated values of both these quantities for the 
$^{41}$K-$^{87}$Rb mixture.


While all the calculations described in the present work are
obtained by solving the two coupled equations \eqref{eq:gpe},
as far as the surface tension and the healing length are concerned, we 
use (as often done in the literature, see for instance Ref.\onlinecite{cabrera2018quantum})
a simpler single-component density functional, 
as briefly described in the following.

The equilibrium density of a droplet at $T=0$
is obtained by requiring the vanishing of
the total pressure, which yields the condition \cite{petrov2015quantum}

\begin{equation}
\dfrac{\rho_2}{\rho_1} = \sqrt{\dfrac{g_{11}}{g_{22}}},
\label{equi}
\end{equation}

If one assumes that this optimal composition is
realized everywhere in the system, the energy functional \eqref{functional}
becomes effectively single-component, and can be written in terms of a single density only.
By defining the following coefficients:
\begin{equation}\label{alpha}
        \alpha  = \frac{1}{4}\left( \frac{\hbar^2}{2m_1}+\frac{\hbar^2}{2m_2}\sqrt{\frac{g_{11}}{g_{22}}}\right)
        \end{equation}
        \begin{equation}\label{beta}
        \beta = g_{11}+ g_{12}\sqrt{\frac{g_{11}}{g_{22}}}
        \end{equation}
        \begin{equation}\label{gamma}
\gamma = \frac{8}{15 \pi^2}\left(\frac{m_1}{\hbar^2}\right)^{3/2}g_{11}^{5/2} \left[1+\left(\frac{m_2}{m_1}\right)^{3/5}\sqrt{\frac{g_{22}}{g_{11}}}\right]^{5/2}
\end{equation}
the effective single-component energy density of the mixture,
expressed for simplicity in terms of the density $\rho _1$ of the first species, reads:
\begin{equation}
    \mathcal{E} = \alpha\frac{(\nabla \rho_1)^2}{\rho_1} + \beta \rho_1^2 + \gamma \rho_1^{5/2}
\label{singlec}
\end{equation}
so that $E=\int d{\bf r} \, \mathcal{E} $.

Self-bound quantum droplets are, by definition, systems with a finite
surface tension. 
Remarkably, the surface tension for a planar
interface separating a self-bound quantum liquid 
from vacuum
can be estimated, without any prior knowledge of the density profile,
by calculating the following integral \cite{stringari_treiner},
\begin{equation}\label{sigma}
    \sigma =2 \int_{0}^{\rho_0} d\rho_1\, \sqrt{\alpha \left(\beta \rho_1+\gamma \rho_1^{3/2}-\mu_0\right)}\, ,
\end{equation}
where $\mu_0=\beta \rho +\gamma \rho ^{3/2}$ is the chemical potential of a liquid system
in equilibrium with the vacuum, evaluated at the equilibrium density $\rho =\rho _0$.

\begin{figure}
    \centering
    \includegraphics[width=1.0\linewidth,clip]{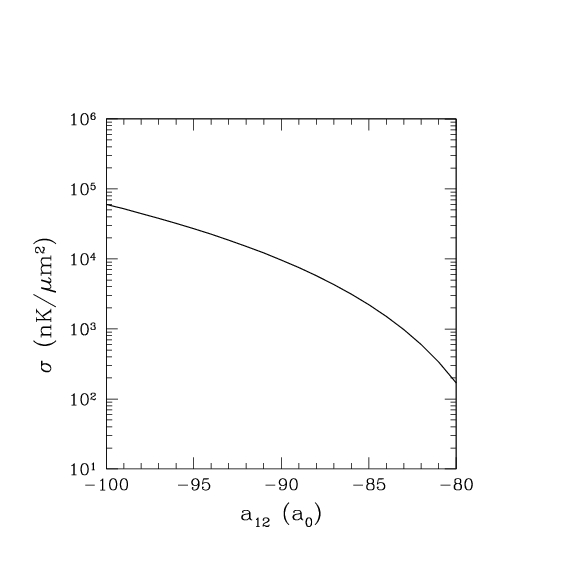}
    \caption{Surface tension of the $^{41}$K-$^{87}$Rb quantum liquid 
as a function of the inter-species scattering length $a_{12}$ (from Ref.\onlinecite{anci_poli}).} 
    \label{fig01}
\end{figure}

The surface tension of the binary mixture $^{41}$K-$^{87}$Rb has been computed 
for different values of the interspecies scattering length $a_{12}$ in Ref. \onlinecite{anci_poli}.
It turns out that relatively
small changes in the interspecies interaction strength cause
order-of-magnitude changes in the surface tension\cite{anci_poli}, which ranges 
from $\sigma \sim 10^2\,nK/\mu m^2$ for $a_{12}=-80\,a_0$ 
to $\sigma \sim 10^5\,nK/\mu m^2$ for $a_{12}=-100\,a_0$.
We show for clarity in Fig. \ref{fig01} the values of $\sigma $ for the $^{41}$K-$^{87}$Rb mixture,
as calculated in Ref. \onlinecite{anci_poli}.

An explicit expression for the healing length 
of the self-bound $^{41}$K-$^{87}$Rb mixture can be obtained
as the length scale where the 
kinetic energy of the system equals the chemical potential.
In this way one can derive
the following expression for the healing length 
in the first species and second species, respectively:

\begin{equation}\label{healing}
\xi _1=\left(-\frac {2\alpha }{2\beta \rho_1+\frac {5}{2}\gamma \rho_1 ^{3/2}}\right)^{1/2}
\end{equation}

\begin{equation}\label{healing1}
\xi _2=\left(-\frac {2\alpha }{2\beta \sqrt{\frac {g_{22}}{g_{11}}} \rho_2
+\frac {5}{2}\gamma  (\frac {g_{22}}{g_{11}})^{3/4}  \rho_2 ^{3/2}}\right)^{1/2}
\end{equation}

\begin{figure}
    \centering
    \includegraphics[width=1.0\linewidth,clip]{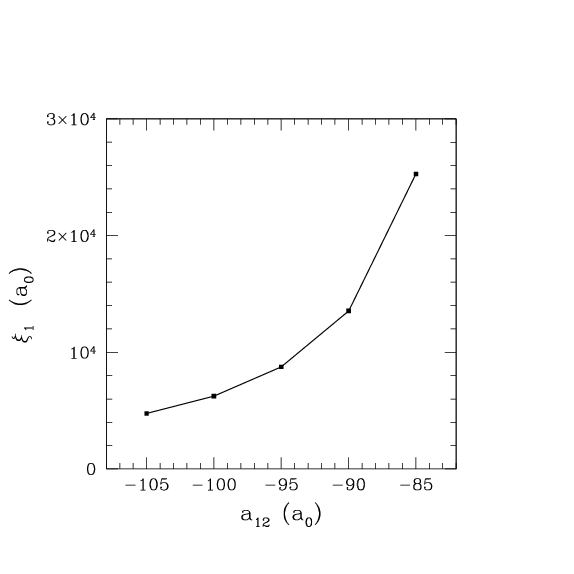}
    \caption{Healing lengths of the $^{41}$K-$^{87}$Rb quantum liquid 
as a function of the inter-species scattering length $a_{12}$.} 
    \label{fig02}
\end{figure}

Notice that, if the optimal ratio between the two densities
is exactly realized, then $\xi _1=\xi_2$. 
We plot in Fig. \eqref{fig02} the calculated healing length $\xi _1$
for different values of the interspecies scattering length $a_{12}$.


\subsection{Vortices in the extended system}

In order to achieve a better understanding
of the intrinsic properties of vortices in the $^{41}$K-$^{87}$Rb mixture, 
we first studied, by solving the two-component system \eqref{eq:gpe},
isolated vortices in an {\it extended} $^{41}$K-$^{87}$Rb system for different
values of the scattering length $a_{12}$ in the range where the 
formation of self-bound liquid is expected: for this specific mixture 
this occurs for $a_{12}<-75.4\,a_0$.

The initial state is represented by 
Eq. \eqref{eq57}, where we take $\alpha=0$ and imprint just a single vortex in the center
of the system, with $\rho _0 $ equal to the bulk density for the 
species hosting the vortex.
The flow field of a linear vortex has a long-range character,
$\sim 1/r$, $r$ being the distance from the vortex axis.
We have imposed, during the minimization, 
antiperiodic boundary conditions \cite{Pi07} in order
to satisfy the condition of no-flow across the boundary of
the computational cell.

A measure of the vortex excitation energy per
unit length of a linear vortex of length $L$ is given
by the integrated vortex kinetic energy \cite{reatto,rossi}, which can be
defined as follows:

\begin{equation}
\epsilon_v(R) =\frac{1}{L} [E_{kin}^v(R)-E_{kin}^0(R)]
\label{vort_kin}
\end{equation}
where $E_{kin}^v,E^0_{kin}$ are the kinetic energies
within a cylinder of radius $R=\sqrt{x^2+y^2}$ and length $L$
(with and without a vortex line along the $z$-axis, respectively)
as a function of the distance $R$
from the vortex line.
$E_{kin}^v(R)$ is given by the integral
$\int _0 ^L dz 2\pi \int _0 ^R dR' \, R' \, \epsilon (R')$,
where $\epsilon (R)= 
 (\hbar ^2/2m)|\nabla_R \psi _i|^2$, and
$\nabla_R \equiv (\partial/\partial x,\partial/\partial y)$
($i$ is the index of the vortex-hosting component of the mixture). 
Similar expressions hold for $E^0_{kin}(R)$.

We notice that the classical hydrodynamical counterpart of $\epsilon _v(R)$
for a vortex in an incompressible fluid of density $\rho _0$ and
circulation $\kappa $ is

\begin{equation}
\epsilon _v ^{hydro}(R)=\frac {\kappa ^2}{4\pi} m
\rho_0 \left[\ln\left(\frac{R}{d_v}\right)+\delta \right]
\label{hydro_class}
\end{equation}
where $d_v $ is the vortex core radius and $\delta$
depends on the model for the core ($\delta=0$
for the hollow core model and $\delta=1/4$ for a
core in rigid rotation) \cite{donnelly}. The parameter $\delta$ in the previous
equation can be absorbed in the logarithmic
term; using the quantum value for the circulation, $\kappa=h/m$,
$\epsilon_v ^{hydro}$ reads

\begin{equation}
\epsilon _v ^{hydro}(R)=\frac{\hbar^2}{m}\pi
\rho _0 \, \ln\left(\frac{R}{\lambda}\right)
\label{hydro}
\end{equation}
for a singly-quantized vortex, where $\lambda =d_ve^{-\delta}$ is the core parameter.

We compute the lowest energy solution, for different values of the
inter-species scattering length $a_{12}$, starting from the
initial state given by Eq. \eqref{eq57}, and then make a best-fit interpolation of
the calculated vortex kinetic energy 
$\epsilon _v(R)$, as defined in Eq. \eqref{vort_kin}, with 
the hydrodynamic approximation for the vortex excitation energy 
(the vortex being in the species $i$) given by Eq. \eqref{hydro},
using $\rho _0$ and $\lambda $ as fitting parameters.
We will consider two cases: (i) both species embed a singly-quantized vortex
with a common core position
and (ii) a singly-quantized vortex is imprinted in one species only.
Therefore in the first case
the phase change around the core position
is $2\pi$ when calculated separately for the two species,
while in the second case the phase change is zero for the species
which does not embed any vorticity.

We find that the hydrodynamic expression \eqref{hydro} accurately 
reproduces the calculated excitation energies far from the vortex line.
By comparing the calculated kinetic energy 
with that predicted by Eq. (\ref{hydro})
we computed the core parameter $\lambda $ as a function of
$a_{12}$.
We found, with good accuracy, that 
$\lambda_i =1.07\,\xi_i$ when both species host a singly-quantized vortex. When
a vortex is imprinted in the first species only, K, we find 
$\lambda_1 =0.82\,\xi_1$, whereas we find $\lambda_2 =0.53\,\xi_2$ for a vortex in the 
Rb species only. These ratios are independent on the chosen values for $a_{12}$. 

\begin{figure}
    \centering
    \includegraphics[width=1.0\linewidth,clip]{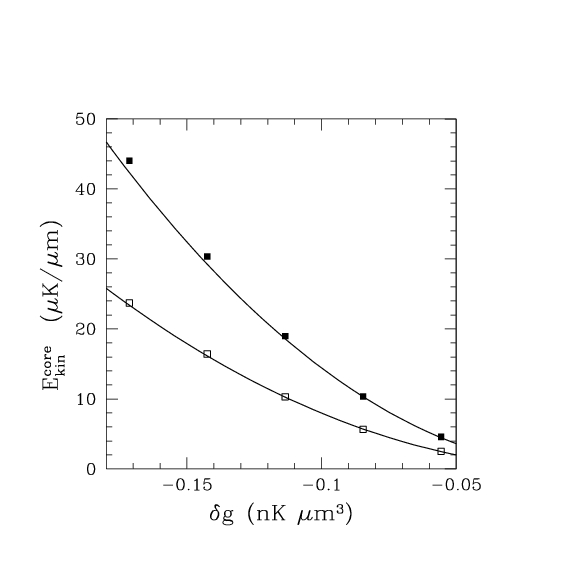}
    \caption{Core energies for the two species as a function of the parameter 
$\delta g$. 
Filled squares: first species (K); open squares: second species (Rb).
The solid lines are obtained from Eq. (\ref{parabolic}).}
    \label{fig1}
\end{figure}

As expected,
the approximation breaks down at distances approaching
the vortex core, as there the local density becomes very
small. The radius $R_c$ of the vortex core region is
defined here as the distance from the center at which the modulus
of the wave function is equal to half the value it
takes far from the core region. From this, the
vortex ``core" energy, defined as $E^{core,i}=E_{kin,i}(R_c)$, is thus obtained.
The calculated values are shown in Fig.(\ref{fig1}) 
as a function of $\delta g=g_{12}+\sqrt{g_{11}g_{22}}$,
for the case where a 
singly-quantized vortex is embedded in each species.

We found that the core energies $E^{core,i}$ are well approximated by the 
expression (solid lines in Fig. \eqref{fig1})
\begin{equation}
\frac {E^{core,i}}{L}=\frac {\pi}{4} \frac {\hbar ^2}{m_i}K_i \delta g ^2
\label{parabolic}
\end{equation}
where $K_i$ are the coefficients relating the densities for the 
uniform system to $\delta g^2$, i.e.
\begin{equation}
\rho_1=\frac {25\pi}{1024}\frac {1}{a_{11}^3}\frac {f^{-2}(z,1,x)}{g_{11}g_{22}}\delta g^2\equiv K_1\delta g^2
\label{rho-deltag}
\end{equation}
and similarly for $\rho_2 =\sqrt{g_{11}/g_{22}}\,\rho_1$.

\begin{figure}
    \centering
    \includegraphics[width=1.0\linewidth,clip]{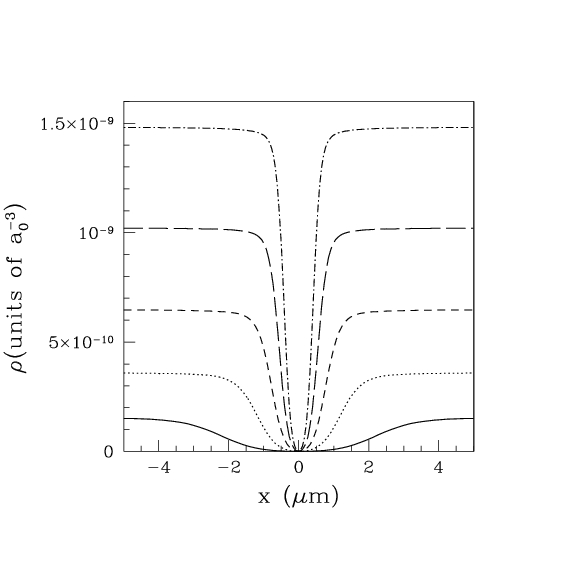}
    \caption{Total density $\rho_1 +\rho_2$ along a line passing through the core of a vortex 
imprinted in both species,
for different values of the interspecies scattering length $a_{12}$.
From top to bottom: $a_{12}=-105,\,-100,\,-95,\,-90,\,-85\,a_0$.
}
    \label{fig2}
\end{figure}

When a singly-quantized vortex is imprinted in both species, the 
resulting density profile is shown in Fig.(\ref{fig2}).
As the scattering length becomes more negative, the vortex core 
shrinks. The core in the total density is empty since both species host a vortex line.

\begin{figure}
    \centering
    \includegraphics[width=1.0\linewidth,clip]{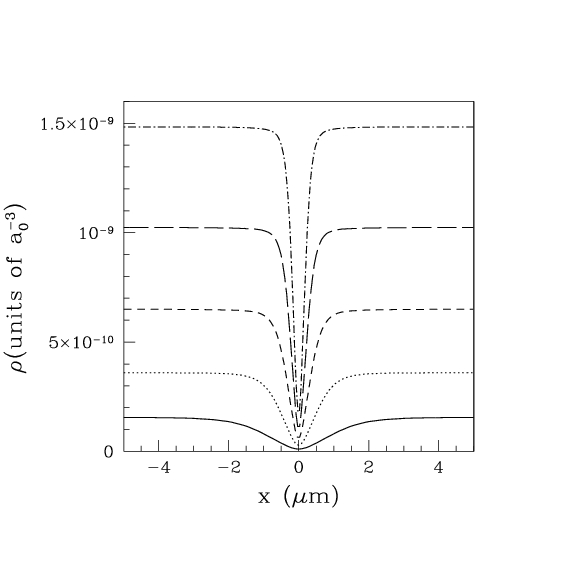}
    \caption{ Total density $\rho_1 +\rho_2$ along a line passing through the core 
when a single vortex is nucleated in the second species (Rb), 
for different values of the interspecies scattering length $a_{12}$.
From top to bottom: $a_{12}=-105,\,-100,\,-95,\,-90,\,-85\,a_0$.}
    \label{fig3}
\end{figure}

When the 
vortex is in the second, heavier species only (Rb), the profiles look like in Fig. (\ref{fig3}).
Notice that the vortex core size is reduced with respect to the previous case, and moreover
the core is partially filled by the first species, which hosts no vorticity.
Similar profiles, characterized by a core that is partially filled with the species without vorticity,
are found when the vortex is imprinted in the first species only (K). 

\begin{figure}
    \centering
    \includegraphics[width=1.0\linewidth,clip]{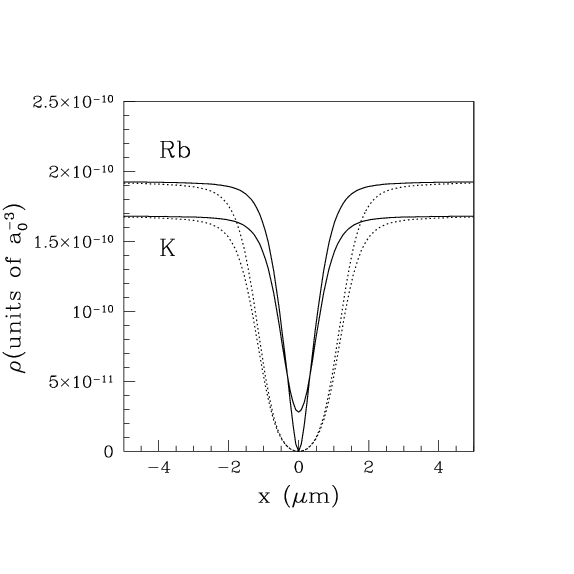}
    \caption{Density profiles for each species, for the case $a_{12} = -90$ $a_0$,
for a vortex imprinted in both species (dotted lines),
and a vortex imprinted only in the second one, Rb (solid lines): notice that, 
in the latter case,
the K density displays a partially filled core, as discussed in the text.}
    \label{fig4}
\end{figure}

We compare in 
Fig. (\ref{fig4}) the vortex structure, close to the core region, 
for the two cases: (i) both species host a singly quantized vortex, and (ii)
only the second species hosts a singly quantized vortex (a similar behaviour is 
found in the case where only the first species carries a vortex).
In spite of the fact that only the second species carries vorticity, a deep depression 
develops also in the vortex-free species 1, mimicking a ``fake" vortex core
with a small residual density at the core position. 
A similar effect has been theoretically predicted for 2-dimensional QDs made of
binary homonuclear Bose mixture \cite{kartashov2020structured}.
This is a consequence of
the fact that the system tries to restore everywhere the optimal 
ratio between the densities (Eq. \eqref{equi}), except very close to the core center.
We find that the amount of filling due to the species without vorticity 
decreases as $a_{12}$ becomes less negative.

\begin{figure}
    \centering
    \includegraphics[width=1.0\linewidth,clip]{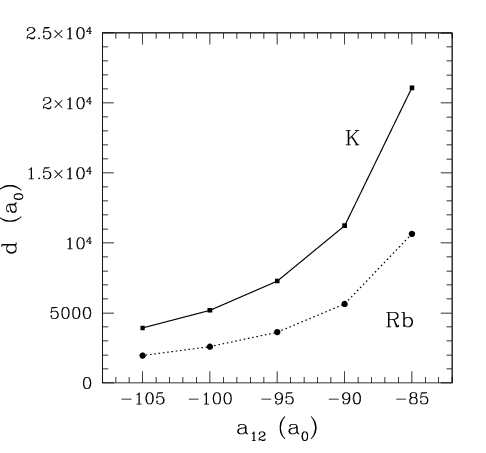}
    \caption{Widths of the vortex cores, defined as the half-width at
half-maximum of the density value far from the vortex, for the 
case of a vortex in the first species (K) only (black squares)
and for a vortex in the second species (Rb) only (black dots). 
}
    \label{fig4bis}
\end{figure}

From the calculated density profiles we computed the 
widths of the vortex cores, defined as the half-width at
half-maximum of the density value far from the vortex position,
as a function of the scattering length $a_{12}$.
We show the results in Fig. \eqref{fig4bis} for the cases where 
a vortex is imprinted either in one species or the other.
It appears that the core widths in the lighter species, K, are twice 
as large as in Rb, and increase with $a_{12}$, as expected from 
the behaviour of the healing length in Fig. \eqref{fig02}.
This could turn out to be a useful information to identify
the vortex-hosting species in experiments, where the two
species can be imaged separately.

In order to determine the most energetically favourable configuration
of vortices in the mixture, 
we computed the energy cost to have (i) a vortex in both species and at the 
same position, (ii) two spatially separated vortices,
both in the first species only, (iii) two spatially separated vortices,
both in the second species only, and (iv) two spatially separated vortices, one in the first species
and the other in the second species. 

\begin{figure}
    \centering
    \includegraphics[width=1.0\linewidth,clip]{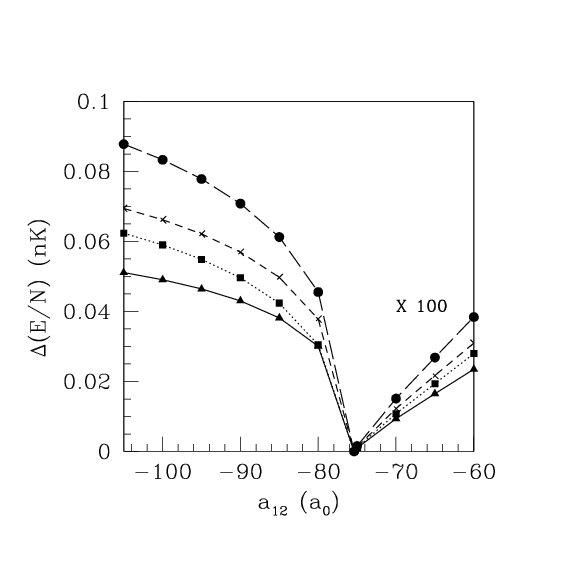}
    \caption{ Vortex formation energies (per pair) as a function of $a_{12}$.
The value $a_{12}=-75.4\,a_0$ marks  
the crossover between the self-bound and the
uniform (unbound) superfluid system. Values to the right of this point are multiplied by a factor 100
for clarity.
Triangles: $2V_2$; squares: $V_{1+2}$; crosses: $V_1+V_2$; dots: $2V_1$.
}
    \label{fig6}
\end{figure}

The results, as a function of interspecies scattering length, 
are shown in Fig. \eqref{fig6}, where we report the calculated 
energies per atom, $\Delta (E/N)$, where:

\begin{equation}
\Delta (E/N)_{2V_i}=2[(E/N)_{V_i}-(E/N)_0]
\end{equation}
for two isolated vortices in the species $i=1$ (K), $i=2$ (Rb);

\begin{equation}
\Delta (E/N)_{V_{1+2}}=(E/N)_{V_{1+2}}-(E/N)_0
\end{equation}
for two vortices in the same position, one inside each species;

\begin{equation}
\Delta (E/N)_{V_1+V_2}=\sum _{i=1}^2 (E/N)_{V_i}-2(E/N)_0
\end{equation}
for two isolated vortices, one in the first species and one in the 
second species.
Here $(E/N)_0$ is the energy for the vortex-free, uniform system.

We remark that in order to compute the energy of two isolated (i.e. spatially separated) 
vortices $V_i$ and $V_j$ 
($i,j=1,2$) we simply add the calculated energies 
(obtained by solving the GP equations \eqref{eq:gpe})
of {\it single} vortex configurations $V_i$,
so it is as if the two vortices were non interacting with one another. 

The results in Fig. \eqref{fig6} show that the most energetically favoured
configuration is the one
with a vortex in the second species (Rb),
whereas the least energetically favoured
is that with vortices in the first species (K) only.

The fact that nucleating vortices in the heavier species is
energetically favoured could be explained by recalling that the vortex
energy \eqref{hydro} is essentially proportional to the ratio $\rho /m$,
which is lower for the Rb species.
 

The same energetic ordering, albeit with much smaller energies (which
appear multiplied by a factor $100$ in Fig. \eqref{fig6} for clarity)
persists also for the unbound superfluid mixture, i.e. 
for $|a_{12}|<75.4\,a_0$.

\subsection{Nucleation of vortices from dynamical simulations}

A possible way of producing in experiments QDs hosting vorticity could be
to first nucleate vortices in a superfluid, un-bound mixture (i.e. with 
$a_{12}$ greater than the critical value for the stabilization of self-bound droplets) 
subject to harmonic confinement by, e.g., 
stirring the condensate with a laser beam
or by using a rotating, slightly
ellipsoidal radial trap. Then the scattering length $a_{12}$ should be quenched to a
more negative value, where the formation of a (much denser) QD is expected, and the harmonic
confinement released at the same time. 
If the size of the droplet is large enough to accomodate one or more vortices,
the final configuration would likely be a vortex-hosting QD.

Although such process could be simulated by performing time-dependent 
numerical simulations based on the extended GP equation described in 
Section II, it implies however a huge computational cost in the present settings,
due to the very large size of the droplets 
needed in order to host few stable vortices, 
and due to the fine mesh in real
space required to accurately represent the wavefunctions, especially in the vortex 
core regions.
Therefore, we consider a
$^{41}$K-$^{87}$Rb mixture 
in a rotating cylindrical trap (described in the following) aligned with the
rotational axis.
We use periodic boundary conditions
along the z direction, where the densities of the two species
are constant (the system is
translationally invariant along this direction). 
In this way 
we reduce the calculations to 
an effectively 2-dimensional sytem. Calculations are
performed in the corotating frame, using Eq. \eqref{eq7} and
with fixed angular velocity.

We will only address a simplified version of the vortex nucleation process here, 
i.e. the dynamical nucleation of vortices in a rotating, trapped mixture in the superfluid,
unbound phase (we notice that the stability of vortex states in a superfluid
binary mixture in two-dimensions has been studied in Ref.\onlinecite{kuopan}). 
This will allow us to verify the above prediction that 
nucleation of singly-quantized vortices in the 
Rb phase only will most likely occur. A more systematic study of the full 
process (i.e. the vortex nucleation in the rotating superfluid phase, followed 
by a quench of the interspecies scattering length into the
self-bound regime, with the likely formation of vortex-carrying quantum droplets) 
will be the subject of a future study.

The interspecies scattering length is set to $a_{12} = -70 \, a_0$
so that the system is just inside the miscible regime with $\delta g > 0$.
Since the system is in the gaseous phase, an additional harmonic confining 
potential is necessary in order to stabilize it. 
The number of atoms for each species is $N_1 \, = 10^6 $ and $ N_2 \, = \, 1.1765 \times 10^6$,
i.e. the atom numbers satisfies the optimal ratio in Eq. \eqref{equi}, $N_1/N_2=\sqrt{g_{22}/g_{11}}$.
We use different trapping potentials acting on each species, through an additional term
in the energy functional (\ref{eq:energyfun}):
\begin{equation}
\mathcal{E}_{ho}[\rho_1,\rho_2]  =  \sum_{i=1}^2 \frac{1}{2} m_i \left(\omega_{i,x}^2 x^2 
+\omega_{i,y}^2  y^2 \right) \rho_i({\bf r})
    \label{harmonic_functional}
\end{equation}
We choose here the trapping 
frequencies in such a way that the two species experience 
the same force constant along each direction, i.e.
$m_1 \omega_{1,\alpha}^2 = m_2 \omega_{2, \alpha}^2 $ ($\alpha = x, y$) 
We also introduce a slight anisotropy in the trapping potential,
$\omega_{1,x}/\omega_{1,y}  = 1.1$,
which favours the nucleation of vortices as the trap is rotated.
The values used are
$(\omega_{1,x}, \omega_{1,y}) = 2 \pi \times (6.50, 5.91) $ Hz and
$ (\omega_{2,x}, \omega_{2,y}) = 2 \pi \times (4.46, 4.06) $ Hz.
As for the rotational frequency in 
the corotating frame (see Eq. \eqref{eq5}) we use the value
$\omega = 2 \pi \times 3.1 $ Hz. The chosen value for $\omega $
must be higher than the critical value necessary to nucleate a single vortex line,
which is of the order of $\omega_c \, = \, \frac{\hbar}{m R^2} \, \ln\left( \frac{R}{\lambda} \right)
\sim 2 \pi \times 0.9 $ Hz,
where $R$ is the average condensate radius in the x-y plane and $\lambda $ is the vortex core parameter.

\begin{figure}
    \centering
    \includegraphics[width=1.0\linewidth,clip]{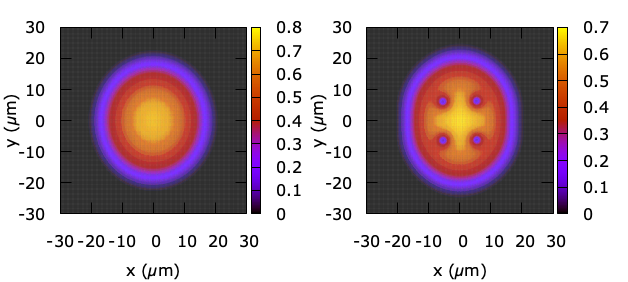}
    \caption{ 2D total density profiles on the $xy$-plane of the SF cylinder. 
On the left the equilibrium density profile at
rest is shown: the deformation is due to the anisotropic harmonic
potential. The right part of the figure 
shows the stationary configuration in the corotating frame 
with angular frequency $\omega$, with 
four vortices nucleated in the second species (Rb). Lengths are expressed in $\mu m$, densities are
expressed in units of $10^{14}$ cm$^{-3}$.}
    \label{fig8}
\end{figure}

The initial configuration of the imaginary-time dynamics is shown
in the left panel of Fig. (\ref{fig8}), which represents the ground-state
configuration in the (stationary) elliptical trap. 

The outcome of the imaginary-time dynamics 
is the spontaneous nucleation of four vortex 
lines in the second species (Rb), which enter the 
cylinder from the lateral surface, 
and then move inside the bulk region until they reach a stationary position. 
In the final configuration, that is shown in the right panel of Fig. (\ref{fig8}),
these vortex lines are located at the same distance from the rotational axis. 
Angular momentum is stored in the first species (K) only through the quadrupolar 
deformation favoured by the elliptical trap. 
We remark that the configurations shown in the figure are stationary  
in the corotating frame;
as a consequence, they would be seen in the laboratory frame
as if they were rotating with the angular frequency $\omega$.
This is indeed what we observed after
performing a real-time dynamics with $\omega =0$ starting from the 
configuration shown in the right panel of Fig. (\ref{fig8}).

\begin{figure}
    \centering
    \includegraphics[width=0.8\linewidth,clip]{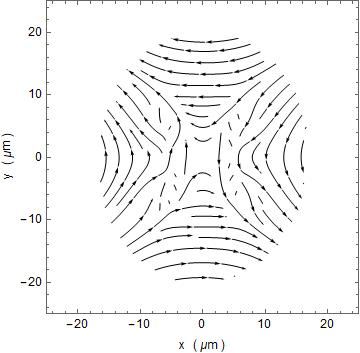}
    \caption{Superfluid flow for the first species (K), shown with streamlines in the $xy$-plane
orthogonal to the rotation axis, for the rotating 
configuration shown in the right part of Fig. (\ref{fig8}).
}
    \label{fig7bis}
\end{figure}

\begin{figure}
    \centering
    \includegraphics[width=0.8\linewidth,clip]{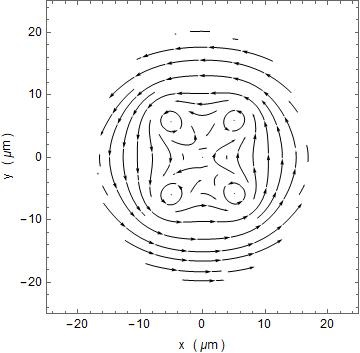}
    \caption{Superfluid flow for the second species (Rb), shown with streamlines in the $xy$-plane
orthogonal to the rotation axis, for the rotating configuration shown in the right part of Fig. (\ref{fig8}).
}
    \label{fig7ter}
\end{figure}

Streamlines of the superfluid flow are shown in Figs. \ref{fig7bis} and \ref{fig7ter},
illustrating the irrotational velocity
fields in each component.
Streamlines allow to infer by visual inspection
the coexistence of vortices and surface capillary waves, as their velocity fields are very different.
The streamlines associated to vortices wrap around their cores, as in  
Fig. \eqref{fig7ter}, 
whereas those associated to capillary waves
end abruptly at the surface of the superfluid \cite{Fet74,Sei94,Oco20},
as in Fig. \eqref{fig7bis}: in the laboratory frame
this results in the rotation of the cloud as a whole, with angular frequency $\omega $.

\subsection{Critical droplet sizes for vortex stability}

Not all values of $N=N_1+N_2$ are
allowed in a self-bound droplet for a given $a_{12}$ 
because small droplets, made with a total number of atoms below 
some critical value $N_c$, become
unstable when the kinetic energy dominates over the
interaction energy, eventually causing the evaporation of the droplet itself.
The critical size $N_c$ for quantum droplets has been calculated 
for the $^{41}$K-$^{87}$Rb mixture in Ref. \onlinecite{anci_poli}.
In the presence of vortices, however, the critical size is expected to be larger,
since the droplet must accomodate the vortex structure and the associated velocity field.
We have estimated such critical size by computing a vortex-hosting 
droplet structure and check its stability during the evolution in time.

We have studied first the case of two singly-quantized vortices
in a $^{41}$K-$^{87}$Rb quantum droplet, one in each species.
In this case three possible outcomes of the imaginary-time evolution 
in the co-rotating frame are found:
(i) {\it unstable} regime: during the evolution in imaginary time
the vortex core is gradually expelled from the droplet, which eventually 
recovers the lowest energy structure of a stable, vortex-free one; 
(ii) {\it metastable} regime: during the evolution in imaginary time 
the system apparently converges towards a stable configuration 
with the vortex in the center of the droplet. However, starting 
a real-time dynamics from this
state the system slowly (in a time of the order of $5$ ms)
expels the vortex and the droplet again recovers a stable, vortex-free structure;
(ii) {\it stable} regime: the system converges towards a stable configuration
with the vortex cores aligned along the $z$-direction and in the center of the droplet.
These configurations are found to be robust against real-time evolution initiated
from this converged stationary state.

\begin{figure}
    \centering
    \includegraphics[width=1.0\linewidth,clip]{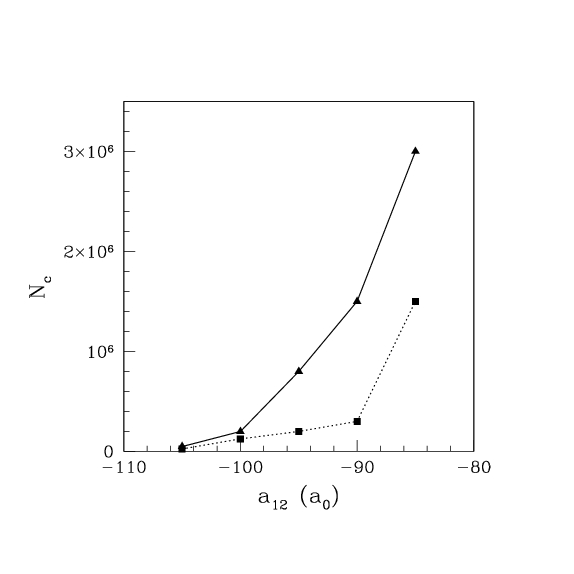}
    \caption{ Stability diagram showing the minimal size of
droplets hosting a vortex. Solid line: one vortex in both species;
dotted line: one vortex in the second species (Rb) only.
}
    \label{fig7}
\end{figure}

In the case of the most energetically favoured 
configuration, {\it i.e.} one vortex in the second species
only, we find however that there is not a metastable region: the vortices are either
unstable and are eventually expelled from the droplet, 
or they stabilize inside the droplet. The
critical line separating stable from unstable vortices is shown in Fig. (\ref{fig7}), compared
with the similar line for the double-vortex case. As it is shown in Fig. \eqref{fig4}, 
the core sizes in this configuration are smaller with respect to the previous one:
this gives the possibility to smaller droplets to sustain a single vortex.

From the comparison with the critical size for vortex-free K-Rb droplets \cite{anci_poli}
it appears that the critical size for stability of a single vortex 
is much larger, of the order of $10^6$ atoms. For comparison, at $a_{12}=-85\,a_0$
the critical size for vortex-free droplets is $N\sim 20,000$.
We must notice that droplets of sizes above the lower critical line in Fig. (\ref{fig7}) 
have not been experimentally realized so far for the K-Rb mixture. 

\subsection{Angular momentum and shapes in rotating K-Rb quantum droplets}

The shapes of classical liquid droplets undergoing rigid-body rotation follow
a universal stability diagram in terms of reduced angular
momentum $\Lambda $ and reduced angular velocity $\Omega $ (defined in the following).
The configurations of rotating droplets lie on two possible branches 
in the $\Omega -\Lambda $ plane \cite{Hei06,But11,Ber17}:
\begin{itemize}
    \item an ascending $\Omega(\Lambda)$ branch for lower values of $\Lambda $ 
corresponding to oblate axisymmetric shapes: the higher $\Lambda $,
the more squeezed is the droplet along the rotational axis;
    \item a descending $\Omega(\Lambda)$ branch for higher values of $\Lambda $ 
describing prolate (i.e. non axisymmetric) shapes, such as
ellipsoids, capsules, and dumbbells: the higher $\Lambda $, the more elongated
are such droplets along an axis perpendicular to the rotational axis.
\end{itemize}
The two branches meet at the point $\Lambda =1.2$ where they form a cusp.

The analysis of superfluid $^4$He droplets rotating solely through
capillary waves shows the presence of an additional descending $\Omega(\Lambda)$ branch in the
stability diagram \cite{Anc18,Anc21}, that is peculiar to superfluid:
this branch is populated by prolate (i.e. non-axisymmetric) droplets, since these are the
only configurations that can store
a finite amount of angular momentum in the form of capillary waves.

We will use here rescaled units, as usually done for classical liquid droplets,
which allow to compare our
results for different droplet sizes and values of $a_{12}$, and also to 
compare our results with the ones for $^4$He rotating droplets,
in spite of the orders-of-magnitudes differences in surface tensions and densities.
Such units, in the case of a binary mixture, are defined as 
\begin{equation}
 \Omega \equiv \sqrt{\frac{(m_1\rho _1+ m_2\rho _2)\, R^3}{8 \, \sigma}}\, \omega
\end{equation}
\begin{equation}
\Lambda \equiv \frac{1}{\sqrt{8 \sigma R^7 (m_1\rho_1+m_2\rho_2)}} \, L_z
\label{eq4}
\end{equation}
where $\sigma $ is the surface tension of the $^{41}$K-$^{87}$Rb 
mixture, shown in Fig. \eqref{fig01}, and
$R$ is the sharp radius of
a (spherical) droplet with $N=N_1+N_2$ atoms, defined such that $4 \pi R^3 (\rho_1+\rho_2)/3 \, = \, N$.

\begin{figure}
    \centering
    \includegraphics[width=1.0\linewidth,clip]{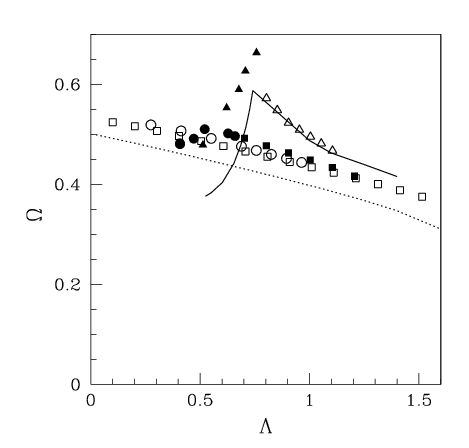}
    \caption{ Rescaled angular velocity $\Omega$ vs. rescaled angular momentum $\Lambda$. 
Solid line: $^4$He droplets\cite{Anc15} hosting vortices ($N=1500$); dotted line: prolate vortex-free $^4$He droplets\cite{Anc15}
($N=1500$);
black triangles: oblate 3-vortex droplets ($N=9\times 10^6$, $a_{12}=-95\,a_0$); 
open triangles: prolate 3-vortex droplets
($N=1.5\times 10^6$, $a_{12}=-105\,a_0$);
open squares: prolate vortex-free droplets ($N=10^5$, $a_{12}=-105\,a_0$);
open circles: prolate vortex-free droplets ($N=10^5$, $a_{12}=-90\,a_0$); 
black squares: prolate vortex-free droplets
($N=1.5\times 10^6$, $a_{12}=-105\,a_0$); black dots: prolate 2-vortex droplets 
($N=9\times 10^6$, $a_{12}=-95\,a_0$).
}
    \label{fig9}
\end{figure}

We investigated different configurations of QDs hosting a finite 
amount of angular momentum,
with and without vortices, although
the fully three-dimensional geometry used here and the need of fine meshes
in real space severely limit the maximum sizes and number 
of vortices that we can address.

\begin{figure}
    \centering
    \includegraphics[width=1.0\linewidth,clip]{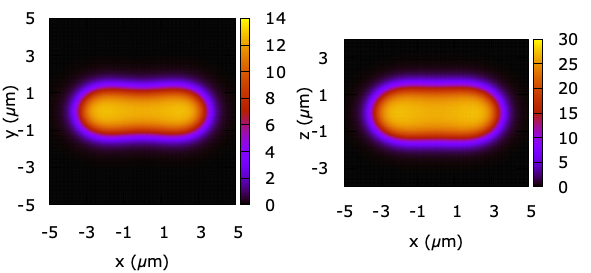}
    \caption{  Left: total density $\rho $ on the $xy$-symmetry 
plane perpendicular to the rotational axis
for a prolate, vortex-free droplet with $N=10^5$; right: side view in the $xz$ plane
passing through the center of the droplet. Lengths are expressed in $\mu m$, densities are
expressed in units of $10^{14}$ cm$^{-3}$.
} 
    \label{fig10}
\end{figure}

The equilibrium shapes of the rotating droplets are strongly influenced by the 
way in which angular momentum is stored (i.e. via capillary waves and/or vortices).
The vortex-free droplets lie, as expected, 
on a single branch which characterizes prolate shapes: 
the angular momentum in these droplets can only be stored in the form of 
capillary waves, as discussed in the Introduction.
The calculated 
points in the $\Omega -\Lambda $ plane corresponding to prolate, vortex-free droplets
(open squares and open circles in Fig. \eqref{fig9})
are very close to the curve found for prolate $^4$He droplets\cite{Anc21}, as shown in Fig. \eqref{fig9}.
This (almost) universal behaviour is a remarkable 
result given the very different nature of these
two quantum liquids, whose surface tensions and densities differ by many orders of
magnitude.
An example of prolate vortex-free droplet (with $N=10^5$ and for $a_{12}=-105\,a_0$)
is shown in Fig. \eqref{fig10}, corresponding 
to a value of the angular momentum $\Lambda = 1.1$. The two plots show the total density 
of the droplet in the $xy$-symmetry plane perpendicular to the rotation axis, and in 
the $xz$ plane containing the rotation axis.

\begin{figure}
    \centering
    \includegraphics[width=1.0\linewidth,clip]{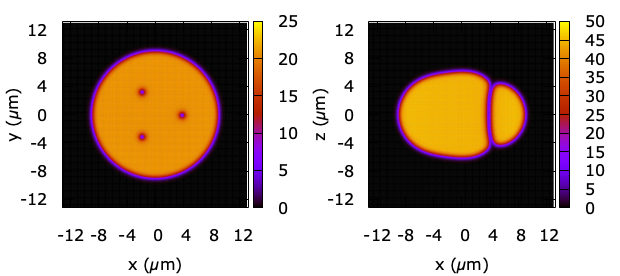}
    \caption{  Left: total density $\rho $ on the $xy$-symmetry
plane perpendicular to the rotational axis
for an oblate droplet hosting three vortices with $N=1.6\times 10^7$; right: side view in the $xz$ plane
passing through the center of the droplet. Lengths are expressed in $\mu m$, densities are
expressed in units of $10^{14}$ cm$^{-3}$.
}
    \label{fig11}
\end{figure}

\begin{figure}
    \centering
    \includegraphics[width=1.0\linewidth,clip]{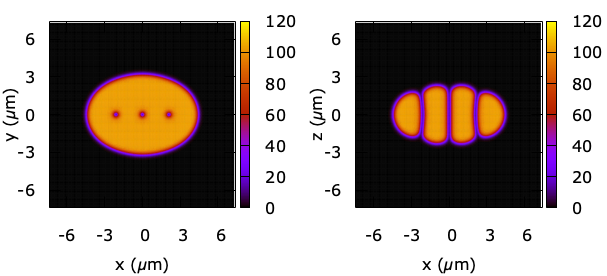}
    \caption{Left: total density $\rho $ on the $xy$-symmetry 
plane perpendicular to the rotational axis
for a prolate droplet hosting three vortices with $N=1.5\times 10^6$; right: side view in the $xz$ plane
passing through the center of the droplet. Lengths are expressed in $\mu m$, densities are
expressed in units of $10^{14}$ cm$^{-3}$.
}
    \label{fig12}
\end{figure}

When vortices are present in the droplet, however, the calculated points seem
to follow a more ``classical" behaviour, characterized by a rising branch 
for oblate, vortex-hosting QDs, and a decreasing branch where prolate, vortex-hosting QDs
lie instead (open and black triangles in Fig. \eqref{fig9}). 
An example of oblate (axi-symmetric) droplet (black triangles in the rising branch in Fig. \eqref{fig9}) 
hosting three vortices in its interior 
(with $N=1.6\times 10^7$ and for $a_{12}=-90\,a_0$), corresponding
to a value of the angular momentum $\Lambda = 0.7$,
is shown in Fig. \eqref{fig11} by means
of equal total density maps in the $xy$ and $xz$ symmetry planes, as in Fig. \eqref{fig10}
(only the vortex whose core is contained in the $xz$ plane passing through the
center of the droplet appears in the right panel of Fig. \eqref{fig11}).

Finally, an example of prolate droplet (open triangles in the decreasing branch in Fig. \eqref{fig9})
hosting three aligned vortices in its interior
(with $N=1.5\times 10^6$ and for $a_{12}=-90\,a_0$)
corresponding
to a value of the angular momentum $\Lambda = 0.9$, 
is shown in Fig. \eqref{fig12}.
At variance with the case of oblate, vortex-hosting droplets, where 
the angular momentum is associated mainly with the vortices, in the case 
of prolate, vortex-hosting droplets like the one shown in Fig. \eqref{fig12}
angular momentum is shared between vortices and capillary waves, the latter
being associated with the loss of axial symmetry in the $xy$ plane 
(in analogy with the case of spinning $^4$He droplets\cite{Anc21}). 

Again, the quasi-classical behaviour of the rotating quantum droplets
with vortices almost matches that of liquid $^4$He nanodroplets, as it appears from Fig. \eqref{fig9}.
Differences however are present, which are most likely due to finite-size effects
and to the small number of vortices (in general, the larger the number of vortices, the 
closer is the behaviour of the spinning droplets to the classical ones\cite{Anc21}).
Unfortunately, we were only able to investigate droplets with 2 or 3 vortices at most,
due to the excessive computational burden discussed before.

\section{Conclusions}

We have studied spinning, self-bound quantum droplets made with the binary Bose
mixture $^{41}$K-$^{87}$Rb.
A preliminary 
analysis of the extended $^{41}$K-$^{87}$Rb system in the quantum liquid regime shows 
that the configurations with vortices inside the
heavier species, Rb, is the most energetically favoured. 
In the presence of a
vortex line in one component alone, the density of the vortex-free species forms an almost empty 
“fake” core on top of the vortical one. 
The increased stability with one
vortex in the second species is confirmed
by studying the rotation of a trapped mixture in the SF regime and subsequent vortex nucleations.

We have thus studied small $^{41}$K-$^{87}$Rb quantum droplets under rotation,
focusing on the interplay between
capillary waves and quantized vortices, that are the two main 
mechanisms with which angular momentum can be stored in spinning
superfluid droplets.

The resulting ($\Omega$,$\Lambda$) phase diagram 
presents strong similarities, despite 
the orders-of-magnitude differences in densities and surface tension,
with the case of rotating superfluid $^{4}$He nanodroplets.
In particular, while prolate vortex-free quantum droplets,
where angular momentum can only be stored in the form of capillary waves, lie on the 
superfluid branch of the diagram, the vortex-hosting droplets show, instead, a
behaviour similar to classical rotating liquid droplets, again in analogy
with the case of superfluid $^4$He. The shapes of vortex-hosting droplets
can be either axi-symmetric (where the angular momentum is stored in 
the form of singly quantized vortex lines) or prolate (where the angular momentum is shared 
between vortices and capillary waves).
Although 
quantum droplets whose sizes are large enough to host vortices 
have not been experimentally realized so far for the K-Rb mixture,
we believe that this study could be helpful for the interpretation of
future experiments where angular momentum can be deposited in quantum droplets 
by, e.g., setting into rotation the mixture in the superfluid state and 
then quenching it into the 
droplet regime by tuning the interspecies scattering length.

\begin{acknowledgments}
We thank Alessia Burchianti, Manuel Barranco and Marti Pi for useful comments.
\end{acknowledgments}

{99}

\end{document}